# A NEW APPROACH TO QUANTUM GRAVITY: A SUMMARY

Sarah B. M. Bell,[1] John P. Cullerne, Bernard M. Diaz[2]
*Department of Computer Science I.Q. Group,*
*The University of Liverpool, Liverpool, L69 7ZF*

## Abstract

Quantum Electrodynamics (QED) has been so successful a theory that it is taken as a model for the production of further quantum theories. However, when the prescription for quantising electromagnetic interactions that so successfully resulted in QED is applied to General Relativity the theory obtained is not renormalizable. We derive a different method of quantising classical electromagnetism which also results in QED. We call the method the *versatile method*. We then apply the versatile method to General Relativity, in particular the Einstein equation which equates a geometrical description derivable from the metric to the energy-momentum-stress tensor, or as we shall call it the matter tensor of the matter field. The method can be applied provided that there is always a reference frame, which may differ with location and time, where the matter tensor can be reduced to a mass density with the other elements zero. We call such matter tensors *simple.* This restriction means that the tensor can be put into one to one correspondence with the Dirac current. When the versatile method of quantising a classical theory is applied to General Relativity the theory that results is renormalizable. It is in fact isomorphic to QED, provided that the temporal and a spatial co-ordinate are exchanged. We discuss the implications of the new theory of which perhaps the most interesting is the possibility that the electromagnetic and gravitational fields might couple in circumstances that could be realized experimentally. We also consider the quantum theory of more general matter tensors.

## 1. Introduction

*Section 1.1.* This account is a summary of a more detailed version by Bell et al. [2000c]. Complex numbers and scalars are given plain type. Vectors and column or row matrices are given italic type. Other matrices and quaternions are given bold type. Maps are given bold italic type. ‡ signifies quaternion conjugation. Bold type $\mathbf{i}_0 = 1$. Bold type $\mathbf{i}_1 = \mathbf{i}, \mathbf{i}_2 = \mathbf{j}, \mathbf{i}_3 = \mathbf{k}$ stand for the quaternion matrices. Suppose we have a quaternion, $\mathbf{Q}$, or four-vector, $Q$, so that $\mathbf{Q} = q_0 \mathbf{i}_0 + q_1 \mathbf{i}_1 + q_2 \mathbf{i}_2 + q_3 \mathbf{i}_3$ or $Q = (q_0, q_1, q_2, q_3)$. We call $q_0$ the temporal part and the rest the spatial part.

*Section 1.2.* A quantum model for the special case of the self-gravitation of a thin spherically symmetric shell of dust, which we shall call the *thin shell model,* has recently been discussed. A classical equation for the total energy governing the evolution of the shell is found [Berezin et al. 1991], [Kuchar 1968], and this equation is the same as occurs for the electromagnetic energy of a charged shell with the mass replacing the charge in the gravitational case and the sign being altered to reflect the fact that the interaction is attractive. A method of quantisation is then chosen [Berezin 1997a], [Berezin 1997b and references therein], [Dolgov and Khriplovich 1997], [Hajicek et al. 1992

---

[1] We would like to thank Thomas Radley for the inspiration of his theory of the spinning universe. We would like to acknowledge the assistance of E.A.E. Bell

and references therein]. We can see our solution as involving a thin shell of matter. The main differences between our discussion here and the previous model are that we derive rather than guess the method of quantising the system and that our method applies to all gravitational fields provided only that the matter tensor, $T_{\mu\nu}$, is simple everywhere.

## 2. The relativistic Bohr atom

*Section 2.1.* We have already shown that there is a version of QED, we call it *the versatile version,* which has exactly the same energy and momentum eigenvalues as the usual version but permits the bispinor to vary as a four-vector [Bell et al. 2000a]. The bispinor may also be located in Minkowski spacetime where the time co-ordinate is imaginary and the space co-ordinates real [Bell et al. 2000b]. This version of the theory allows a classical interpretation with the bispinor a position vector of a particle which is itself scalar. We show that the relativistic Bohr atom, that is, the circular orbits as treated by Sommerfeld [Born 1962], can be derived from the versatile Dirac and photon equation and vice versa. The notation is taken from Bell et al. [2000a&b]. We commence by deriving *the Proto-Real transformation.* Let $Q = (i\, q_0, q_1, q_2, q_3)$, $q_\mu$ real, be a four-vector field in Minkowski spacetime co-ordinated by $(i\, x_0, x_1, x_2, x_3)$, $x_\mu$ real. We may map $Q$ onto the components of a quaternion $\mathbf{Q}$,

2.1.A $\quad \mathbf{Q} = i\,\mathbf{i}_0\, q_0 + \mathbf{i}_1\, q_1 + \mathbf{i}_2\, q_2 + \mathbf{i}_3\, q_3.$

We may express the elements $q_1$ and $q_2$ in terms of polar co-ordinates $(r, \theta)$ for the plane containing axes one and two,

2.1.B $\quad q_1 = q_s \cos\theta - q_r \sin\theta, \quad q_2 = q_s \sin\theta + q_r \cos\theta,$

where $\delta s = r\, \delta\theta$. We may transform the quaternion matrices to polar co-ordinates by

2.1.C $\quad \mathbf{i}_1 = \mathbf{i}_1 \cos\theta - \mathbf{i}_2 \sin\theta, \quad \mathbf{i}_2 = \mathbf{i}_1 \sin\theta + \mathbf{i}^r_2 \cos\theta,$

from which we find

2.1.D $\quad \mathbf{Q} = i\,\mathbf{i}_0\, q_0 + \mathbf{i}_1\, q_s + \mathbf{i}_2\, q_r + \mathbf{i}_3\, q_3.$

We see that the form of $\mathbf{Q}$ has not changed. Let us also represent our transformation as active with $(i\, x_0, x_1, x_2, x_3)$ as a position vector which suffers a global transformation. We may suppose that lines of equal $x_1$ are curled round to form arcs, s, on the circumference of circles and lines of equal $x_2$ re-orientated to form radii, r. We may specify the scaling applied to the inverse of this transformation as

2.1.A $\quad \delta s = (R/r)\, \delta x_1 \quad\quad \delta r = \delta x_2,$

where R is the value of r for which the interval $\delta x_1$ is neither shrunk nor stretched. We call equations 2.1.A the Proto-Real transformation.

*Section 2.2.* We show that we can use the Proto-Real transformation to provide a curved spacetime. The versatile Dirac equation for an electron is

2.2.A $\quad (\underline{\mathbf{D}} - i\,e\,\underline{\mathbf{A}}^\sim)\, \underline{\Phi} = \underline{\Phi}\, \mathbf{M},$

where we write $\underline{\mathbf{A}}^{\sim}$ for the original $\underline{\mathbf{A}}$ of Bell et al. [2000a&b] and e is the magnitude of the electric charge on the electron. We call the observer with this system of Cartesian co-ordinates *Large observer one*. We now employ the Proto-Real transformation. We transform axes $x_1$ and $x_2$ to polar co-ordinates r and θ with

2.2.B     $s = R\theta, \quad x_1 = -r\sin(s/R), \quad x_2 = r\cos(s/R),$

where R is constant. We set ( $q_s = \{R/r\}\{(\partial/\partial s)_r\}$ ) and ( $q_r = (\partial/\partial r)_s$ ). We transform the quaternion matrices $\mathbf{i}_1$ and $\mathbf{i}_2$ to lie along the newly-defined axes s and r using section 2.1. The transformed Dirac equation is

2.2.C     $(\underline{\mathbf{D}}^B - ie\underline{\mathbf{A}}^{B\sim\prime})\underline{\mathbf{\Phi}}^{B\prime} = \underline{\mathbf{\Phi}}^{B\prime}\underline{\mathbf{M}}^B.$

We call the observer with this system of co-ordinates *Small observer one*. Because $\mathbf{i}_1$ now lies along a curved axis, the space on which the Dirac equation 2.2.C is defined may be seen as curved. If we suppose that the Small observer does not see the volume element, $\delta r\, \delta s\, \delta x_3$, changing with r, he sees his space as a flat Cartesian one with the first and second spatial Cartesian axes Rθ and r where R is constant, but the potential must be altered. Suppose the original volume element for the Large observer is δV and the volume element for the Small observer is $\delta V^B$. We have

2.2.D     $\delta V = \delta x_1\, \delta x_2\, \delta x_3 \rightarrow \delta V^B = \delta r\, \delta s\, \delta x_3,$

where δr is taken at constant θ and δs at constant r. Then from equation 2.1.A we obtain

2.2.E     $\delta V^B = (R/r)\, \delta V.$

We call this conversion of r into R the *Real Transformation*. We discuss the reason for the name in section 4.4. Proceeding as in Bell et al. [2000a&b] we may find the photon equation in terms of the Dirac current which provides the probability current. It is

2.2.F     $\underline{\mathbf{D}}\,\underline{\mathbf{D}}\,\underline{\mathbf{A}}^{\sim} = \underline{\mathbf{J}}^{\sim},$

where we write $\underline{\mathbf{J}}^{\sim}$ for the original $\underline{\mathbf{J}}$ in Bell et al. [2000a&b]. The only property of the current of interest here is that for a suitably chosen frame the temporal current, $J' = J^{\sim\prime} i$, may be expressed as a probability, P, per unit volume element, δv", with the other components of the current zero. We transform equation 2.2.F to this frame using a Lorentz transformation we will call Z. We obtain $J' = P/\delta v"$. Performing the Proto-Real transformation on equation 2.2.F indicated by an obvious notation, we obtain

2.2.G     $\underline{\mathbf{D}}^B\,\underline{\mathbf{D}}^{B\ddagger}\,a^{B\prime} = P/\delta V, \quad \underline{\mathbf{D}}^B\,\underline{\mathbf{D}}^{B\ddagger}\,a^B = P/\delta V^B,$

where for the first equation the potential is $a^{B\prime} = a^{B\sim\prime}i$ and is equal to that seen by the Large observer and for the second the potential is $a^B = a^{B\sim}i$ and is equal to the potential seen by the Small observer. Since the volume elements are constant from the point of view of the appropriate observer we obtain

2.2.H     $\underline{\mathbf{D}}^B\,\underline{\mathbf{D}}^{B\ddagger}\,(a^{B\prime}\,\delta V) = P, \quad \underline{\mathbf{D}}^B\,\underline{\mathbf{D}}^{B\ddagger}\,(a^B\,\delta V^B) = P.$

From equations 2.2.H and using equation 2.2.E to eliminate constants of integration we obtain

2.2.I  $a^B = a^{B'} (r/R)$.

We transform $a^{B\sim'}$ and $\tilde{a}^{B\sim}$ with the inverse of Z to $\underline{A}^{B\sim'}$ and $\underline{A}^{B\sim}$ and the final version of the Dirac equation for the Small observer is

2.2.J  $(\underline{D}^B - ie\underline{A}^{B\sim})\underline{\Phi}^B = \underline{\Phi}^B \underline{M}^B$,

where $\underline{\Phi}^B$ is the solution associated with the new potential $\underline{A}^{B\sim}$.

*Section 2.3.* We derive Bohr's first equation. We discuss the hydrogen atom with an infinitely heavy nucleus at rest. We find that we can solve the Dirac equation from the point of view of the Large observer by considering the solution of the Dirac equation 2.2.J, which uses the Small observer's co-ordinate system, if we choose appropriate boundary conditions. We may find $\underline{A}^{B\sim}$ by using the Real transformation and observing that $\underline{A}^{\sim}$ and $\underline{A}^{B\sim}$ agree when r = R. We may therefore solve the versatile photon equation in its original form, equation 2.2.F, as for a point charge. We obtain, using equation 2.2.I,

2.3.A  $A^{B\sim} = ie/R$,

where $A^{B\sim}$ is the potential term in Minkowski spacetime. We assume the electron has a velocity in the s direction, that is, the electron is circling the proton, and solve equation 2.2.J for the wave function $\underline{\Phi}^B$ when r = R. We discover that

2.3.B  $\{ \phi_1^B = \exp\{i(\nu x_0^{\sim} + \mu s)\}, \phi_2^B = \{(i\nu - i\mathbf{i}_1\mu - ieA^{B\sim})/M^{-1}\}$
$\{$
$\{ \qquad \exp\{i(\nu x_0^{\sim} + \mu s)\}, \tilde{m}^2 = (\nu - eA^{B\sim})^2 + \mu^2,$

where $\phi_1^B$ and $\phi_2^B$ are the bispinors, we omit h, Planck's constant, $\tilde{m} = (m_e/i)$ with $m_e$ the rest mass of the electron and we have a plane wave solution where $(-i\nu)$ is the frequency, $\mu$ the wave number and $x_0^{\sim} = x_0/i$. We see there is an equivalent free electron with wave number $\mu$ and frequency $\eta = \nu - eA^{B\sim}$. This electron can be seen as the one originally considered by Bohr in his theory. We have shown that there is no force on the Bohr electron, but that its circular motion is due to the curvature of its spacetime. We employ the Real transformation using de Broglie's relations for the free electron at distance R from the proton and the condition that the wave function $\underline{\Phi}^B$ must be single valued to provide the relativistic version of the first of Bohr's equations

2.3.C  $\tilde{m}\,\tilde{v}\,R/\sqrt{(1+\tilde{v}^2)} = n(h/2\pi)$,

where $(\tilde{v}/i)$ is the velocity of the electron, n is an integer and we have restored Planck's constant, h.

*Section 2.4.* We derive Bohr's second equation. Given the Dirac equation 2.2.A for a free electron travelling along the $x_1$ axis with frequency $\eta$ measured along the imaginary axis $x_0^{\sim}$ and wave number $\mu$ along $x_1$, we may calculate the frequency $\eta''$ and wave number $\mu''$ along any other pair of axes ($x_0''$, $x_1''$), geometrically, using the dot product of the two four-vectors:

2.4.A  $\eta\,\delta x_0^{\sim} + \mu\,\delta x_1 = \eta''\,\delta x_0^{\sim''} + \mu''\,\delta x_1''$.

Applied to the Bohr electron and hydrogen atom this gives,

2.4.B $\quad \eta \, \delta x_0\tilde{} + \mu \, \delta s = \nu \, \delta x_0\tilde{}$.

This, together with de Broglie's expression for the frequency of the free electron gives $\nu = \tilde{m} \sqrt{(1 + v\tilde{}^{\,-2})}$, and hence, applying the Real transformation

2.4.C $\quad i \, e^2 / R = \tilde{m} \, v\tilde{}^{\,-2} / \sqrt{(1 + v\tilde{}^{\,-2})}$,

the second of Bohr's equations. We may check the validity of equation 2.4.C. The equation follows from the Newtonian equation of circular motion provided that we associate the factor of $\sqrt{(1 + v\tilde{}^{\,-2})}$ with a Special Relativistic increase in mass. In addition, we will treat the equation of circular motion using General Relativity in section 3.2. It is possible to obtain equation 2.4.C from this by equating the mass multiplied by the acceleration four-vector to the appropriate measure of the electric field [Feynman et al. 1964].

*Section 2.5.* The spin angular momentum of the system as discussed above is an integer corresponding to the orbital angular momentum of the Bohr electron. We suppose that this is made up from an orbital angular momentum obtained from solving equation 2.2.A with no Proto-Real transformation, a spin of a half for the electron and another half integer spin attributable to the Berry phase [Anandan 1992]. We locate a Dirac bispinor in Minkowski spacetime [Bell et al. 2000b] and interpret it as the position vector of the Bohr electron from the point of view of Small observer one. Large observer one sees a superposition of all possible orbits of the Bohr electron unless the orbit is pinned by a measurement of angular momentum. Small observer one sees the wave function spread evenly over a disc.

*Section 2.6.* We show the converse: that QED may be derived from the Bohr atom. Suppose we have an infinitesimal sphere of radius R', centered at a point L with a quasi-particle Å of charge f at L. The potential, A, at a point b on the boundary of the sphere is

2.6.A $\quad A = f / R'$.

We may replace $A^\#$ by a constant charge density $\rho$ which we may also consider a probability density as Dirac did. We then have

2.6.B $\quad A = (4/3) \pi R'^2 \rho$.

The d'Alembertian reduces to $(\square = d^2 / dR'^2)$, and so

2.6.C $\quad \square \, A = (8/3) \pi \rho$.

This, apart from a constant indicating a different choice of units, is the photon equation in the rest frame of the current. We may therefore describe a field as producing the charge density at L. We may write down the Dirac equation, 2.2.A, for particle $B^\#$ of charge e and bare mass M' at b.

2.6.D $\quad (\underline{D} - e \underline{A}(\{f/R'\}, \{f/R'\})) \underline{\Phi} = (-i) \underline{\Phi} \, \underline{M}(M', -M')$.

From equation 2.6.A we may substitute for $(f/R')$ obtaining

2.6.E $\quad (\underline{D} - e \underline{A}(A, A)) \, \underline{\Phi}(\phi_1, \phi_2) = (-i) \, \underline{\Phi}(\phi_1, \phi_2) \, \underline{M}(M', -M')$,

so again we may replace $A^{\#}$ by the field described by equation 2.6.C. We suppose that Bohr's equations hold for particle $A^{\#}$ at L and $B^{\#}$ at b, with R' the Bohr radius. From equations 2.3.C and 2.4.C we may calculate the velocity, v', of particle $B^{\#}$, v' = $2\pi$ e f / ( n h ), and a condition on M' and R'

2.6.F   M' R' = { $h^2$ } √{ 1 - $4\pi^2$ $e^2$ $f^2$ / ( $n^2$ $h^2$ ) } / { $4\pi^2$ e f },

where we have replaced the second e in the original equation 2.4.C with f. We may assemble a wave function, $\underline{\Phi}^B$, from equations 2.3.B. We subject the Dirac equation 2.6.D to the Proto-Real transformation so that $\underline{\Phi}^B$ fits. We may then undo this step by the operating with the inverse of the Proto-Real transformation, sending $\underline{\Phi}^B \to \underline{\Phi}$ and regaining equation 2.6.D. So we have justified writing equation 2.6.D. We now have a Dirac equation, 2.6.E, that holds for particle $B^{\#}$ at b as well as a photon equation, 2.6.C, that holds for the field at b. We treat equations 2.6.A, 2.6.B and 2.6.F as simultaneous. Substituting from equation 2.6.A for f and then equation 2.6.B for R' into 2.6.F we obtain

2.6.G   ρ = { $3\pi$ $A^2$ $e^2$ / ( 2 $n^2$ $h^2$ ) } { A ± √[ $A^2$ + ( 4 $M'^2$ / $e^2$ ) ] }.

The behaviour of this is satisfactory. We write equations 2.6.C and 2.6.E above in invariant form and apply a Lorentz transformation Z(b) to them that transforms A into the correct local four-vector potential. We see that if the Bohr equations for particle $B^{\#}$ and $A^{\#}$ hold at every point in the local rest frame, then particle $A^{\#}$ may be replaced by a field obeying the photon equation and particle $B^{\#}$ obeys the Dirac equation since we have constructed a solution. This is the versatile method of quantisation. Sommerfeld's elliptical orbits are included in that they can be treated as circular orbits with a different coupling constant as we will discuss elsewhere. We note that if we substitute ( 1/2 - A ) for A in the above and negate both charges, e and f, we may also deduce the photon and Dirac equations given particles $A^{\#}$ and $B^{\#}$.

## 3.   The Bohr equations for quantum gravity

*Section 3.1.* We show that the gravitational interaction between a point particle that curves spacetime according to the Proto-Quantum metric and a second point particle obeys the Bohr equations 2.3.C and 2.4.C, forming a gravitational equivalent of the hydrogen atom. Since we want to distinguish the gravitational from the electromagnetic equivalent we will call the former *the Thalesium atom*. We will call the equivalent of the Bohr electron for a Thalesium atom *a Geotron*. For Large observer one solving the Dirac equation in flat spacetime we have metric

3.1.A   $ds^2$ = $g_{00}$ $dx_0^2$ + $g_{11}$ $dx_1^2$ + $g_{22}$ $dx_2^2$ + $g_{33}$ $dx_3^2$,

where for a space-like interval

3.1.B   $g_{00}$ = - 1, $g_{11}$ = $g_{22}$ = $g_{33}$ = 1.

We identify a Large observer two who has the same viewpoint as Large observer one except that he is not moving along $x_0$ and a Small observer two, who also does not move along $x_0$, and whose viewpoint is the same as that of the Geotron. For Small observer two we must write the curvature of equations 2.1.A in terms of a curved metric in the variables r and θ describing his spacetime. We require that our metric should lead us to the Bohr equations discussed in section 2. We set

3.1.C   √$g_{11}$ $dx_1$ = √a √r dθ,   √$g_{22}$ $dx_2$ = ( 1 / √a ) √r dr.

By substituting equations 3.1.C into equation 3.1.A we perform the Proto-Real transformation,

3.1.D   $d\tau^2 = a r d\theta^2 + (1/a) r dr^2 + \ldots,$   $d\tau^2 = (a/r) ds^2 + (r/a) dr^2 + \ldots,$

where we apply the equation $s = r\theta$ to Small observer two. We call this metric *the Proto-Quantum metric*. s is Small observer two's time co-ordinate.

*Section 3.2.* We derive Bohr's first equation. For a bound system such as the Thalesium atom, we will require a steady state. We consider a circuit of the Geotron round the centre of attraction in the plane r, $\theta$ where r is constant. From the point of view of Small observer two the Geotron's position vector, $\mathbf{V} = (V^r, V^\theta)^T$, is constant and the co-ordinate system of the Thalesium atom rotates. So we consider parallel transport of the position vector. Following Martin [1995] we have

3.2.A   $\{ \Gamma_\theta = \begin{pmatrix} 0 & -a^2/(2r) \\ 1/(2r) & 0 \end{pmatrix} \}.$

We also have

3.2.B   $d\theta / d\tau = 1/\sqrt{(ra)},$

and

3.2.C   $d\mathbf{V}/d\tau = \{(d\theta/d\tau)\Gamma_\theta\} \mathbf{V}.$

Solving equation 3.2.C we obtain

3.2.D   $V^r = \sin\{(1/2)\sqrt{(a/r)}(\tau/r)\},$   $V^\theta = (-1/a)\cos\{(1/2)\sqrt{(a/r)}(\tau/r)\},$

where we have omitted constants of integration. After n circuits defined according to Small observer two we have

3.2.E   $(1/2)\sqrt{(a/r)}(n\tau/r) = 2\pi,$

We may now state our steady-state quantum condition as

3.2.F   $m\tau = 2\pi r,$

where m is an integer. We may turn this condition into a prescription for the Large observer, re-writing equation 3.2.F as

3.2.G   $m s / r = 2\pi.$

We see this is the condition for Large observer one to see the position vector of the Geotron return to the same orientation after m revolutions. We can therefore apply the Real transformation to equations 3.2.E and 3.2.G setting r to R, the Bohr radius, in both and s to $\tau$ for both observers to agree. We then have,

3.2.H   $\sqrt{(a/2R)} = \sqrt{2} m/n,$

where we choose the positive root. If wanted the negative root we could set $\tau = -s$ instead of choosing the negative root explicitly. We will call this *the negative Real transformation* and setting $\tau = s$ *the positive Real transformation.*

*Section 3.3.* We derive Bohr's second equation. The Lagrangian associated with the second of equations 3.1.D when the rest of the metric is zero is

3.3.A $\quad 1 = (a/r)(ds/d\tau)^2 + (r/a)(dr/d\tau)^2.$

Since s is ignorable, $(a/r)(ds/d\tau) = \gamma$, where $\gamma$ is constant. Hence

3.3.B $\quad -a/(2r^2) = d^2r/d\tau^2.$

We want to consider spacetime as flat and find an expression for the equation of circular motion. The covariant derivative provides

3.3.C $\quad v^\mu = Dx^\mu/Ds = (dx^\mu/ds) + (dx^0/ds)\,\Gamma^\mu_{\alpha 0}\,x^\alpha,$

where the $\Gamma^\mu_{\alpha 0}$ are Christoffel symbols. The geodesic equation for the rotating frame provides $Dv^\mu/Ds = 0$, which becomes

3.3.D $\quad dv^\mu/ds + (dx^0/ds)\,\Gamma^\mu_{\alpha 0}\,v^\alpha = 0.$

From equations 3.3.C and 3.3.D in the plane of rotation and setting the radial component of motion to zero, we have

3.3.E $\quad d^2r/ds^2 = -v^2/r,$

where $v^\mu$ has become v. We now use the positive and negative Real transformations at the Bohr radius, R, to relate equation 3.3.E to equation 3.3.B by setting $\tau$ to s and then $\tau$ to $-s$. This is legitimate providing we impose the quantum condition 3.2.H and its negative transformation analogue. We obtain

3.3.F $\quad v^2 = a/(2R), \quad v^2 = -a/(2R).$

Substituting for v from the second of equations 3.3.F into equation 3.3.E we obtain

3.3.G $\quad d^2r/ds^2 = a/(2R^2).$

The positive Real transformation leads to an attractive force for both the Small and Large observers. The negative Real transformation leads to an attractive force for the Small observer and a repulsive one for the Large observer.

*Section 3.4.* We compare the electromagnetic and gravitational versions of Bohr's equations. We re-state Bohr's equations for electromagnetism

3.4.A $\quad m_e v_e r_e / \sqrt{(1 - v_e^2)} = n(h/2\pi), \quad e_e^2/r_e = m_e v_e^2/\sqrt{(1 - v_e^2)},$

where $e_e$ is the charge on the electron, $v_e$ its velocity and $r_e$ the Bohr radius for Small observer one. We re-state the first of equations 3.3.F as

3.4.B $\quad a / ( 2 r_g'' ) = v_g^2$,

in an obvious notation and set

3.4.C $\quad r_g = r_g'' \sqrt{( 1 - v_g^2 )}$.

Then, using equations 3.4.B, 3.4.C and 3.2.H the gravitational versions of Bohr's equations become

3.4.D $\quad m_{gv} v_g r_g / \sqrt{( 1 - v_g^2 )} = n h / ( 2 \pi ), \quad e_g^2 / r_g = m_{gv} v_g^2 / \sqrt{( 1 - v_g^2 )}$,

where $m_{gv} = m_g$ is the rest mass of a particle feeling the effect of the source field and

3.4.E $\quad e_g^2 = m_g a / 2, \quad 2 \pi e_g^2 c / h = \sqrt{2} m$.

We see that the gravitational version of Bohr's equations is now completely parallel to the electromagnetic. If we set both masses involved in the gravitational case to Planck's mass we obtain

3.4.F $\quad m_g = \sqrt{\{ h c / ( \sqrt{2} \pi G ) \}}$,

where we introduce the speed of light, c, and we have set m = 1 for a Geotron which we assume has a Planck mass. To find the equations for the repulsive case we send $v_g \to i v_g$ in the equations for the attractive case above. Equations 3.4.D mean that as well as the point of view of the Large and Small observer two where the Geotron moves along s but does not move along $x_0$ we can assume the point of view of the Large and Small observers one taken in section 2 where the Geotron also moves along $x_0$.

*Section 3.5.* We find the energy eigenvalues for the gravitational case. We sum the kinetic and potential energies of the Geotron in the frame of Large observer one

3.5.A $\quad E'_n = ( m_g / \sqrt{( 1 - v_g^2 )} ) - ( m_g a / ( 2 r_g ) )$.

Through this classical expression for the energy, we see the connection with the thin shell model [Kuchar 1968]. We suggest, as we did for the hydrogen atom, that for the same appropriate circumstances the orbits of the Geotron will be a superposition of all possible for the Thalesium atom, leading to the Geotron being at rest, and that the stationary Geotron will also be spread through a superposition of all possible positions, exactly as if it were a thin shell. Equation 3.5.A, through equations 3.4.D gives the quantised energy of the Thalesium atom as

3.5.B $\quad E'_n = m_g \sqrt{( 1 - ( 2 m^2 / n^2 ) )}$.

This is a similar expression to some of Berezin's suggested schemes of quantisation for the thin shell model [Berezin 1997a], [Berezin 1997b]. We now consider four points of view. (1.1) is that of equation 3.5.B and Large observer one in which the Geotron is moving along both $x_0$, his temporal axis, and s, a spatial circle for the plane $x_1$, $x_2$ in the metric of equations 3.1.A and 3.1.B. For the next point of view, (1.2), s is temporal and $x_0$ is spatial. So in moving from point of view (1.1) to point of view (1.2) we have swapped a spatial, s, and a temporal, $x_0$, co-ordinate. The versatile Dirac and photon / graviton equations also allow tachyon solutions which may be reached from the usual solution by interchanging a spatial, $x_1$, and temporal, $x_0$, axis. Equation 3.1.B becomes

3.5.C    $g_{00} = 1, \ g_{11} = -1, \ g_{22} = g_{33} = 1.$

The metric of Large observer two is given by equation 3.1.A once the changes given in equation 3.5.C have been applied. Equations 3.1.C become

3.5.D    $\sqrt{g_{11}} \ dx_1 = i \sqrt{a} \sqrt{r} \ d\theta, \quad \sqrt{g_{22}} \ dx_2 = (1/\sqrt{a}) \sqrt{r} \ dr.$

We may perform the Proto-Real transformation by substituting from equation 3.5.D into equation 3.1.A to obtain for the metric of Small observer two,

3.5.E    $d\tau^2 = -a r \ d\theta^2 + (r/a) \ dr^2 \ldots.$

We add one possible choice of metric for the other two axes giving

3.5.F    $d\tau^2 = -a r \ d\theta^2 + (r/a) \ dr^2 + r^2 \ d\zeta^2 + r^2 \sin^2(\zeta) \ d\xi^2,$

where r labels a sphere of area $4\pi r^2$ and $\zeta$ and $\xi$ describe colatitude and longitude. If we append the last two terms of this metric to the first of metrics 3.1.D sections 3.2.and 3.3 would proceed to exactly the same conclusions. From the last of the metrics 3.1.D we obtain

3.5.G    $d\tau^2 = -(a/r) \ ds^2 + (r/a) \ dr^2 + r^2 \ d\zeta^2 + r^2 \sin^2(\zeta) \ d\xi^2.$

We have considered orbits confined to the first two terms of the metric. Purely spatial orbits are possible and lead to similar dynamics as we will discuss elsewhere. An alternative to swapping a temporal and spatial axis is to solve the Dirac and photon / graviton equation in the usual way and then swap the energy and momentum eigenvalues. We then obtain for the energy

3.5.H    $E_n" = m_g v_g / \sqrt{(1 - v_g^2)},$

where $v_g$ is now the inverse of that in equation 3.5.A. We move to point of view ( 2.1 ) in which s is a spatial co-ordinate for the plane $x_1, x_2$ of Large observer one, and for which the Geotron is a tachyon. From this point of view the Geotron is not travelling along $x_0$. From the point of view of observer ( 2.1 ), the observer with viewpoint ( 1.2 ) has a velocity of $v_g$ which has increased the energy by a factor $(1/\sqrt{(1 - v_g^2)})$. This is the factor we added in going from $r_g"$ to $r_g$ in section 3.4. Thus from point of view ( 2.1 ) we obtain the final energy, using equations 3.4.D

3.5.I    $E_n = m_g v_g = m_g \sqrt{2} \ m / n.$

From the point of view (2.2) and of Small observer two (or the Geotron) the path along "s" is simply his normal movement along the time axis as time passes and the Geotron is stationary. For him, the energy due to motion in equations 3.5.I is seen as gravitational in origin and we may calculate on this assumption, using equation 3.2.H, that it is of the same form as for equation 3.5.I. It is
$\gamma \propto \sqrt{(a/r)} \propto 2 \ m / n.$

4.    **Quantum Gravitational Dynamics**

*Section 4.1.* We prove that the equations of QED are valid for Quantum Gravitational Dynamics (QGD). We may provide a connection between the Einstein equation and the photon and Dirac equation by arranging that the matter tensor in the rest frame is equal to the Dirac current in the rest frame. We consider a gravitational field for which the matter tensor is simple everywhere and pick a point L( $x_\mu$ ). We use a Lorentz transformation Z( $x_\mu$ ) to go to the frame where the matter tensor at L has only the $T_{00}$ element non-zero. We pick an infinitesimal spherical volume centred on L for which the matter tensor is constant. We now pick a point at b on the boundary of the volume and consider the effect of the infinitesimal volume on a particle, $B^\#$, at b. The volume acts as though all the matter was concentrated at L and hence as though there was a point particle at L, of mass equal to that contained in the spherical volume. We call the effective particle at L particle $A^\#$. Particle $A^\#$ will generate a metric at b and we choose the boundary conditions so that this is the Proto-Quantum metric. We then quantise the gravitational field produced by $A^\#$ applying the condition in equation 3.2.H. Bohr's equations, 3.4.D, now apply. We adopt the same strategy for every time and location applying to the field. We now have a two-body interaction between particle $B^\#$ and particle $A^\#$ representing the field, with the interaction obeying the infinitesimal form of the Bohr equations at every point in spacetime. This means that the field can be seen as obeying the photon equation while particle $B^\#$ obeys the Dirac equation. We scale our equations at b so that the potential in the local rest frame is equal to that of the original gravitational field, transform our equations by the inverse of Z( $x_\mu$ ) and repeat this for every point in the field. The photon / graviton and Dirac equations now represent exactly the same field as the Einstein equation combined with a point particle interacting with the field. The solution to the photon / graviton and Dirac equations gives us the potential, $A_v$, locally at a point addressed by $x_\mu$ and we may find a Lorentz transformation, Z( $x_\mu$ ), to take $A_v$ to a scalar A. We then have from sections 2.6, 3.4 and above in this section

4.1.A $\quad A = - a( x_\mu ) / ( 2\, r( x_\mu ) )$,

where a( $x_\mu$ ) is the mass and r( $x_\mu$ ) the radius of the infinitesimal spherical volume corresponding to particle $A^\#$ at $x_\mu$. This gives us the local value of E and F in the Proto-Quantum metric tensor given by

4.1.B $\quad d\tau^2 = E\, dt^2 - F\, dr^2 - r^2\, d\zeta^2 - r^2 \sin^2( \zeta )\, d\xi^2$,

where t is time, r labels a sphere of area $4\pi r^2$ and $\zeta$ and $\xi$ describe colatitude and longitude. They are

4.1.C $\quad E( x_\mu ) = a( x_\mu ) / ( r( x_\mu ) ) = - 2\, A, \quad F( x_\mu ) = 1 / E( x_\mu ) = - 1 / ( 2\, A )$,

Having obtained the metric tensor, $g_{\mu\nu}$, we may transform by the inverse of Z( $x_\mu$ ). We may also set ( $2 A = 1 - a / r$ ). This also leads to the photon and Dirac equations. If we then set

4.1.D $\quad E( x_\mu ) = 1 - a( x_\mu ) / ( r( x_\mu ) ) = - 2\, A, \quad F( x_\mu ) = 1 / E( x_\mu ) = - 1 / ( 2\, A )$,

we recognise the familiar Schwarzschild metric for a point source in equation 4.1.B.

*Section 4.2.* We discuss the universe as a quantum state. For a state corresponding to the vacuum, the metric given in equation 3.5.E provides a = r. Substituting in the metric of equation 3.5.E, we may obtain

4.2.A $\quad d\tau^2 = - r^2\, d\theta^2 + dr^2 - r^2\, d\zeta^2 - r^2 \sin^2( \zeta )\, d\xi^2$,

where we have provided another possible choice of metric for the other two axes and take r as the temporal co-ordinate. We may set $dx = d\theta$, $dy = d\zeta$, $dz = \sin( \zeta )\, d\xi$. We see at once that we are looking at a Robertson-Walker metric [Martin 1995] for a flat expanding universe with spatial co-ordinates x, y, z. However, this implies that the universe is continuously gaining mass. Classically the universe is only undetermined at the initial singularity. Suppose that from the point of view of Large observer one, temporal axis $x_0$, a Geotron is added to the universe regularly in each interval $\delta x_0$ at this singularity. From the point of view of an observer inside the universe, say Small observer three, the total amount of matter in the universe is still created at the initial singularity. However, while Large observer one is passing through successive moments along his time, Small observer three is jumping from one of the possible parallel universes, in Deutsch's terminology [Deutsch 1997], to another, but one so nearly identical, if the mass increase is small, that we may suppose that Small observer three matches the memories as being the consecutive history through time that classical General Relativity demands.

*Section 4.3.* We derive the classical limit of QGD showing that it is the usual form of the Schwarzschild metric. We may rewrite the Proto-Quantum metric for the universe, equation 4.2.A, using equation 3.2.H, obtaining

4.3.A  $d\tau^2 = ( 4 m_u^2 / n_u^2 ) dt^{\#2} - ( n_u^2 / 4 m_u^2 ) dr^{\#2} + r^{\#2} d\zeta^2 + r^{\#2} \sin^2(\zeta) d\xi^2,$

where $m_u$ and $n_u$ are the integers m and n in equation 3.2.H, we have a space like interval with $r^\#$ the temporal co-ordinate and $t^\#$ spatial as in section 4.2. We see that the vacuum is given by $2 m_u = n_u$ where we identify $m_u^2$ as proportional to the mass of the universe and $n_u^2$, the square of its level of excitation, as proportional to its radius, $r_g" = r^\#$. Going back to our model of the universe, we deduce the potential due to the universe in empty space, $P_u = n_u^2 / ( 4 m_u^2 ) = 1$. We add a source of mass, ( $a_s / 2$ ), proportional to an integer $m_s^2( a_s )$ and determine the potential at a distance r from the source, with $n_s$ the level of excitation and $n_s^2$ proportional to r. The potential due to the source is, from equation 3.2.H, $P_s = - 4 m_s^2 / n_s^2$. The source by itself would give rise to the metric

4.3.B  $d\tau^2 = P_s dt^2 - ( 1 / P_s ) dr^2 + r^2 d\zeta^2 + r^2 \sin^2(\zeta) d\xi^2,$

where t is the temporal co-ordinate. However, any third particle in the universe feels the potential not only of the source but also of the universe. So we take the total potential to be

4.3.C  $P = P_u + P_s.$

As we discussed in sections 2.6 and 4.1 we can quantise this potential equally well. Substituting P for $P_s$ in equation 4.3.B and using equation 4.3.C we obtain

4.3.D  $d\tau^2 = \{ 1 - ( 4 m_s^2 / n_s^2 ) \} dt^2 - \{ 1 - ( 4 m_s^2 / n_s^2 ) \}^{-1} dr^2 - r^2 d\zeta^2 - r^2 \sin^2(\zeta) d\xi^2,$

and we see the usual Schwarzschild solution. As long as $n_s$ is large P can vary smoothly and the system will behave classically. As $n_s$ becomes small the jumps between different Bohr orbits will become apparent, and the system will be overtly quantised. We suppose that a more complicated but still simple system can be modelled by considering $m_s$ and $n_s$ to be functions of the spatial and temporal co-ordinates and then applying the same principle.

*Section 4.4.* We may look at the difference between an attractive and repulsive interaction from the point of view of the Large observer as due to the different type of spacetime manufactured by the interaction. Let the metric be $g_{rs}$, r, s = 1 or 2. For an attractive interaction we have the non-zero elements in the metric tensor as $g_{22} = g_{11}$, while for the second we take them as $g'_{22} = - g'_{11}$. If we take a hyperbolic trajectory for the position vector associated with an orbiting particle for the first, then sending $g_{rs} \rightarrow g'_{rs}$ will transform the trajectory into a circle as discussed by Bell et al. [2000a]. So instead of holding that the two trajectories are different, one a hyperbola and the other a circle we could instead hold that both orbits are circular but apply to different spacetimes. However, we can unify these two spacetimes. We may swap $ds^2$ and $d\tau^2$ if we are going to use the Real transformation and re-write metric 3.5.G as

4.4.A $\quad d\tau^2 = (r/a) ds^2 + (r/a)^2 dr^2 + (r^3/a) d\zeta^2 + (r^3/a) \sin^2(\zeta) d\xi^2.$

We see that the spacetime signature has changed. For the universe on the other hand we may set $r = a$, obtaining

4.4.B $\quad d\tau^2 = ds^2 + dr^2 + r^2 d\zeta^2 + r^2 \sin^2(\zeta) d\xi^2.$

*Section 4.5.* We explain both QED and QGD in a unified way as due to the curvature of spacetime and suggest the gravitational and electromagnetic fields may couple. For the excitation of the electron or Geotron in response to absorbing or emitting a photon or graviton we may described the atom as a superposition of all Bohr orbits. However, a measurement of the angular momentum entails a transition from a superposition of all possible Bohr orbits into one in particular. Also, this particular orbit depends on the orientation chosen for the measurement. The transition from a superposition of all possible orbits into one in particular is where the non-linear equations, the Einstein equation and the equation of motion derivable from the Lagrangian, become the linear photon / graviton and Dirac equations. Our remarks of section 3.1 and 3.2 would still hold if instead we had used the following metric for Small observer two

4.5.A $\quad d\tau^2 = -(a/r) dt^2 + (r/a) dr^2 + dx_3^2 + dx_0^2,$

which describes the curvature as restricted to a single plane. We suppose this is the primary state in the absence of any interaction. We suggest that when two point particles interact the plane containing the curvature is the same for both. Conversely, if they do not interact then it is different. We explore the general case for curvature generated by a field interacting with another particle in detail. From section 2 it suffices to consider the interactions as taking place between particles $A^\#$ and $B^\#$ in infinitesimal volumes throughout the field. We take a point of view of where we are Large observer one for the electromagnetic spin and Large observer two for the gravitational spin. We take as metric equation 3.1.A. We suppose that the spin of the particles has two components, the gravitational component that always occupies the plain $x_0$, $x_r$, the point of view of Large observer two, and the electromagnetic component that always lies in a plane $x_s$, $x_r$, $x_s \neq x_r$, the point of view of Large observer one, where the line joining the particles is along $x_r$. Thus gravitational mass and electromagnetic charge generate two different curvatures in two different planes which couple separately. If either particle or both is neutral then we can describe the situation by having the electromagnetic spin for $A^\#$ in the planes, for example, $x_s$, $x_r$ or $x_r$, $x_q$, $x_s \neq x_q$, $x_q \neq x_r$, while the electromagnetic spin for $B^\#$ lies in the plane $x_s$, $x_q$, where $A^\#$ and $B^\#$ "choose" different planes. If the size of the system and the field of view of the observation are similar then the spin of the system can be orientated as described. If not we suggest this switches on curvature for all orientations for both the gravitational and electromagnetic spins, leading to metrics 3.5.F and 3.5.G. In this case the gravitational parts of the spin will lie in the planes $x_0$, $x_r$, and $x_0$, $x_s$, and $x_0$, $x_q$, while the electromagnetic part will lie in the planes, for example, $x_s$, $x_r$ and $x_s$, $x_q$, and $x_s$, $x_0$. The two will couple in the plane $x_s$, $x_0$. We call this phenomenon *electro-gravity*. Experiment is called for here since there may be alternative scenarios to the one we have described, but there are some indications that the effect has been observed [Modanese and Schnurer 1998 and references therein], [Seife 1999 and references therein], [Podkletnov and Nieminen 1992]. We suggest that their superconducting rotating disc contains electrons which form a large coherent electromagnetic source that can be described as a single quantum state with a single wave function and that must be excited or de-excited as a whole. We call the source a quasi electron. The quasi electron acts like a Bohr electron with the protons in the disc (which do not form a coherent state) as nucleus. Then when we test the quasi electron using the Geotrons in a small neutral body we should find that "gravitational" effects are present. Detailed, calculations are to be found in Bell et al. [2000c] and these describe the interaction qualitatively. The strength of the electro-gravity coupling would have to be determined experimentally.

## 5. Beyond simple matter tensors

*Section 5.1.* We discuss more complicated matter tensors. We may generalise the Dirac and the graviton equations of QGD to two indices. The Dirac current, $J_\nu$, becomes a tensor, $J_{\mu\nu}$, equivalent to the matter tensor, $T_{\mu\nu}$. However we are not deriving the two index equations from the two-body solution which we explored previously and their validity is hence a guess. We may consider a tensor, $J_{\mu\nu}$, to be the Cartesian product of the elements of two four-vectors, $Q_\mu$, $U_\nu$, $J_{\mu\nu} = Q_\mu U_\nu$. This means that we may represent a tensor by $\mathbf{J}_{\mu\nu} = \sum\sum J\tilde{\phantom{}}_{\mu\nu} \mathbf{i}_{\mu\nu}$, where $\mathbf{i}_{\mu\nu} = \mathbf{i}_\mu \mathbf{i}_\nu\hat{\phantom{}}$. $\mathbf{i}_\nu\hat{\phantom{}}$ is a second copy of the quaternion matrices that commutes with the first and ~ signifies that the temporal component of $Q_\mu$ and $U_\nu$ should be multiplied by i. We call $\mathbf{J}_{\mu\nu}$ a *quaternion double*. We allow the reflector and rotator matrices defined by Bell et al. [2000a&b] to have quaternion double elements, and define

5.1.A  $\quad\{\quad \underline{\mathbf{J}} = \underline{\mathbf{J}}(\ \sum\sum J\tilde{\phantom{}}_{\mu\nu} \mathbf{i}_{\mu\nu},\ \sum\sum J\tilde{\phantom{}}_{\mu\nu}" \mathbf{i}_{\mu\nu}^{\ddagger}\ ),$
$\quad\quad\quad\{$
$\quad\quad\quad\{\quad \mathbf{R}_S| = \mathbf{R}_S|(\ \sum\sum R_S\tilde{\phantom{}}_{\mu\nu} \mathbf{i}_{\mu\nu},\ \sum\sum R_S\tilde{\phantom{}}_{\mu\nu} \mathbf{i}_{\mu\nu}\ ),$
$\quad\quad\quad\{$
$\quad\quad\quad\{\quad \mathbf{R}_T| = \mathbf{R}_T|(\ \sum\sum R_T\tilde{\phantom{}}_{\mu\nu} \mathbf{i}_{\mu\nu},\ \sum\sum R_T\tilde{\phantom{}}_{\mu\nu} \mathbf{i}_{\mu\nu}^{\ddagger}\ ).$

Rotations and Lorentz transformations of $\underline{\mathbf{J}}$ can now be transformed into multiplications by $\mathbf{R}_S|$ and $\mathbf{R}_T|$ in the same way as discussed by Bell et al. [2000a]. This means that we can prove that the Dirac and graviton equations augmented by a second index are invariant. The equations are

5.1.B  $\quad (\underline{\mathbf{D}}_{\mu\nu} - i e \underline{\mathbf{A}}_{\mu\nu}\tilde{\phantom{}}) \underline{\mathbf{\Phi}}_{\mu\nu} = \underline{\mathbf{\Phi}}_{\mu\nu} \underline{\mathbf{M}}_{\mu\nu}, \quad \underline{\mathbf{D}}_{\mu\nu} \underline{\mathbf{D}}_{\mu\nu} \underline{\mathbf{A}}_{\mu\nu}\tilde{\phantom{}} = \underline{\mathbf{J}}_{\mu\nu}\tilde{\phantom{}}.$

*Section 5.2.* Let $v$ be the covariant velocity four-vector. For velocity v,

5.2.A  $\quad v = (\ 1, v_1, v_2, v_3\ ) / \sqrt{(\ 1 - v^2\ )}.$

We then have $J_\nu = \rho v_\nu$ and $J_{\mu\nu} = v_\mu J_\nu$, where $J_{\mu\nu}$ applies to equations 5.1.B and $J_\nu$ applies to equations 2.2.A and 2.2.F. We see that solutions to the latter are a subset of the former.

*Section 5.3.* We adopt the usual summation conditions where appropriate. We show that equations 5.1.B are compatible with weak gravity. For this, Martin [1995], the graviton equation with a source term is

5.3.A  $\quad 8\pi T_{\mu\nu} = (\ -1/2\ ) \mathbb{1}\ A_{\mu\nu},$

where

5.3.B  $\quad A^\rho_\sigma \equiv h^\rho_\sigma - (\ 1/2\ ) \chi^\rho_\sigma h^\alpha_\alpha, \quad g_{\mu\nu} = \chi_{\mu\nu} + h_{\mu\nu}, \quad \chi_{\mu\nu} = \text{diag}\ (\ 1, -1, -1, -1\ ),$

$g_{\mu\nu}$ is the metric tensor and terms of second order have been discarded. The graviton equation, 5.3.A, has contraction operations instead of quaternion double multiplications. We may replace the former with the latter arriving at the second of equations 5.1.B. We also have $\partial_\alpha T^\alpha_\nu = 0$. This holds for the quaternion double form of the free Dirac equation, the first of equations 5.1.B, and approximately for small potentials. Introducing equation 5.1.B instead removes the approximation.

## 6. Conclusions

Finally, our conclusion is that if we accept all the implications of the photon, Dirac and Einstein equations for simple matter tensors a quantum theory similar to QGD as described here is likely. We also flag the possibility that the quantum gravitational matter tensor may be simple in all cases. The non-simple matter tensor might be associated with either classical fields like those of Maxwell's electrodynamics, which should be replaced by QED, or may be occasioned by averaging over small volumes in the same way as the matter tensor describing the pressure of a gas may be defined by averaging over particle speed and direction in a tensor composed of particle velocity and density.

---

[3] A fuller account of this paper with algebraic working and more description